\documentclass[fleqn,twoside]{article}
\usepackage{espcrc2}
\usepackage{epsfig}

\usepackage{graphicx}
\usepackage[figuresright]{rotating}

\newcommand{\AmS}{{\protect\the\textfont2
  A\kern-.1667em\lower.5ex\hbox{M}\kern-.125emS}}

\title{Status of APEmille}
\author{A. Bartoloni\address[OLDC]{INFN, Sezione di Roma I, Italy},
	P. Boucaud\address[OLDB]{LPT, University of Paris Sud, Orsay, France},	
	N. Cabibbo\addressmark[OLDC],
	F. Calvayrac\address[OLDE]{LPEC, Universit\'e du Maine, Le Mans, France},
	M. Della Morte\address[OLDMI]{Physics Dept.,
	University of Milano Bicocca and INFN, Sezione di Milano, Italy}$^{,}$%
	\address[OLDK]{DESY/NIC, Zeuthen, Germany},
	R. De Pietri\address[OLDG]{Physics Dept., University of Parma
	and INFN, gruppo collegato di Parma, Italy},
	P. De Riso\address[OLDH]{Physics Dept., University of Roma Tor Vergata
	and INFN, Sezione di Roma II, Italy},
	F.~Di Carlo\addressmark[OLDC],
	F.~Di Renzo\addressmark[OLDG],
	W.~Errico\address[OLDI]{INFN, Sezione di Pisa, Italy},
	R.~Frezzotti\addressmark[OLDMI],
	T.~Giorgino\addressmark[OLDK],
        J.~Heitger\address[NEWA]{Universit\"at M\"unster, Germany},
	A.~Lonardo\addressmark[OLDC],
	M.~Loukianov\addressmark[OLDK],
	G.~Magazz\'u\addressmark[OLDI],
	J.~Micheli\addressmark[OLDB],
	V.~Morenas\address[OLDL]{LPC, Universit\'e Blaise Pascal and IN2P3, Clermont, France},
	N.~Paschedag\addressmark[OLDK],
	O.~P\'ene\addressmark[OLDB],
	R.~Petronzio\addressmark[OLDH],
	D.~Pleiter\addressmark[OLDK],
	F.~Rapuano\addressmark[OLDC],
        J.~Rolf\address[NEWB]{Humboldt Universit\"at, Berlin, Germany},
	D.~Rossetti\addressmark[OLDC],
	L.~Sartori\addressmark[OLDI],
	H.~Simma\addressmark[OLDK],
	F.~Schifano\addressmark[OLDI],
	M.~Torelli\addressmark[OLDC],
	R.~Tripiccione\address{Physics Dept., University of Ferrara and INFN,
	Sezione di Ferrara, Italy},
	P.~Vicini\addressmark[OLDC],
	P.~Wegner\addressmark[OLDK]}

\begin{document}

\begin{abstract}
This paper presents the status of the APEmille project, which is
essentially completed, as far as machine development and construction is
concerned. Several large installations of APEmille are in use for physics
production runs leading to many new results presented at this conference.
This paper briefly summarizes the APEmille architecture, reviews the status
of the installations and presents some performance figures for physics codes.
\end{abstract}

\maketitle

\section{OVERVIEW}
The APEmille project was started with the goal of developing and commissioning 
a massively parallel computer optimized for lattice gauge theory (LGT),
with a peak performance close to the 1 TFlops threshold \cite{Lat99}. 
APEmille is the third generation of APE machines \cite{FirstAPE}. 
The project is now completed, as far as
hardware development and construction are concerned: several APEmille systems 
have been installed at various sites, providing an overall processing power close
to 1.5 TFlops. These systems have become the main workhorse for doing
LGT simulations for several research groups \cite{CCP}.
In this paper, we briefly recall the features of APEmille, 
present the status of the project, describe the final shape of the software 
environment and discuss the performance achieved by physics programs.

\section{APEmille ARCHITECTURE}
APEmille is based on a three dimensional mesh of nodes connected by a synchronous
communication network linking first neighbours. All nodes in the mesh operate
in Single Instruction Multiple Data (SIMD) mode. Each node consists of a 
0.5 GFlops custom developed floating point processor and 32 MB data memory. 
At each clock cycle the processor is able to complete the operation $a \times b + c$
following the IEEE standard, where $a,b,c$ are single precision complex or double
precision real operands.
APEmille is a highly modular and scalable system.
The basic building block and smallest independently running entity is
a processing board (PB) with $2\times 2\times 2$ nodes. 
Up to 16 PBs are interconnected via a
single backplane to form an APEmille crate. One crate has 128 processors,
approximately 65 GFlops peak performance and 4 GB data memory.
Larger systems, which can be re-partitioned by software,  are assembled by 
connecting $n$ crates together. The corresponding topology is $2n \times 8 \times 8$.

\begin{figure}[t]
\vspace*{0.10cm}
\begin{center}
\includegraphics[scale=0.24]{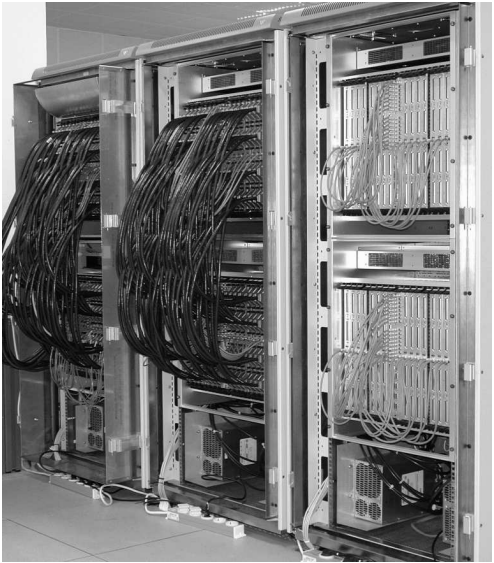}
\end{center}
\vspace*{-1cm}
\caption{\it Three APEmille racks.}
\vspace*{-0.7cm}
\end{figure}


\begin{table}[t]
\caption{\it Key parameters of APEmille:}
\begin{tabular}{ll}
\hline
Peak performance  & 528 MFlops/proc\\
Clock frequency   & 66 MHz \\
FP registers      & 512 (32-bit)\\
Data memory       & 32 MByte/proc \\
Communication BW  & 66 MByte/s/direction \\
I/O BW per master & 6 MByte/s \\
Power consumption & 28 W/GFlops \\
Price             & 2.5 Euro/MFlops peak\\
\hline
\end{tabular}
\vspace*{-0.5cm}
\end{table}


APEmille systems are connected to a network of host PCs with a Linux
operating system, each hosting 4 PBs. The network contains one or
more master PCs on which users log in to start their application
programs. Input/output is usually performed onto disks belonging to the 
master. This I/O setup is limited by the performance of the network connecting
the PCs (typically FastEthernet). The measured bandwidth is of the order of 6 
MBytes/sec. Higher I/O
rates can be achieved by hooking 'local' disks directly to the host PCs with 
Ultra2 SCSI channels. In this setup the bandwidth scales with the size 
of the system. Typically it is of the order of 100 MBytes/sec
per crate.

Power consumption of APEmille systems is very low (less than 30 W/GFlops) and the
footprint of a two-crate rack is about $0.7 m^2$. For these reasons, APEmille
machines are simply air cooled and do not need complex infrastructure.

\section{PROJECT STATUS}

After prototypes were assembled and tested in late 1999, a first round
of production was carried out in the year 2000. A second and final round
of production was started in spring 2001.
APEmille systems have been installed at several sites belonging to the
APEmille collaboration and also at a few more universities and research labs
(see Tab.~1).
The presently installed peak processing power is about 1460 GFlops and will
grow to about 2.2 TFlops once the final round of construction is completed.


APEmille is a very stable system: up-times of 1-2 months or 
more are routinely achieved.
Hardware maintenance is typically limited to simple 
replacement of ageing modules and is therefore cheap, both in terms of hardware 
costs and manpower.

\begin{table}[t]
\caption{APEmille installations by the end of 2001. For each site the total
peak performance is given. Sites in italic refer to institutions
not belonging to the APE collaboration.}
\label{table:1}
\vspace*{1mm}
\begin{tabular}{@{}lr}
\hline
Site             & GFlops \\
\hline
Rome I           & 650 \\
DESY/NIC Zeuthen & 540 \\
Pisa             & 260 \\
Rome II          & 260 \\
Milano           & 130 \\
Bari             & 65  \\
Uni. Paris Sud   & 16  \\
\hline
{\it Bielefeld}  & 145 \\
{\it Swansea}    &  80 \\
{\it INFN-LNGS}  &  65 \\
\hline
Total		 & 2211\\
\hline
\end{tabular}
\vspace*{-0.3cm}
\end{table}

\section{SOFTWARE AND PERFORMANCE}
APEmille systems use the familiar TAO programming language, already
used in all previous APE machines. The language has been extended by
very few elements to exploit the new features of APEmille,
like double precision and local integer data types, and the increased
number of registers.
Old APE100 programs can be re-compiled for the new machine with almost no
changes. High performance, however,  is achieved only after some tuning of
the codes. To this end the TAO programmer still has to follow only a few 
optimisation guidelines.

Typically, the efficiency is limited by latencies in local memory accesses 
and by both latencies and bandwidth in remote data transfers. 
Good performance requires therefore some care to 
hide the latencies and to have data already available in registers
when they are needed by the floating-point unit.
 
The steps to boost efficiency include:
\begin{itemize}
\item Intensive use of registers to prefetch data.
\item Memory accesses which are known to be always local are 
flagged to the compiler, so that a more aggressive scheduling 
can be employed.
\item Memory accesses are made in large bursts of data
(say up to 36--96 complex numbers in a single burst) and 
complex conjugation is performed "on the fly" when loading the data.
\item Use of APEmille intrinsic 64-bit floating-point format
      in precision critical parts of the code.
\end{itemize}   

Using the steps described above to optimize a bi-conjugate 
gradient solver for the improved Wilson-Dirac operator, the following 
performance numbers have been obtained:
\begin{itemize}
\item The main loop in the most time consuming kernel of the code achieves
      a pipeline filling of more than 80\%.
\item Computations with $SU(3)$ matrices (clover term) run at 380 MFlops
      per node.
\item The full inverter runs at 200 MFlops sustained (distributed
      lattice with local volume $6 \times 3 \times 3 \times64$ on each node).
\end{itemize}

\section{CONCLUSIONS}
With the completion of the last APEmille systems at various sites
by end of this year, an other 2 TFlops of overall compute power 
will be available to the LGT community. APEmille has become a main 
and reliable workhorse for many groups. 
System and compiler software are very stable, but further adjustments 
may bring additional improvements in performance and user-friendliness.
While the APEmille project has basically been finished, the 
development of a next generation of APE machines is in progress 
\cite{Lele}.

\section*{ACKNOWLEDGMENTS}
We would like to thank F.~Aglietti, M.~Cosimi, I.~D'Auria, A.~Hoferichter,
A.~Menchikov, A.~Michelotti, E.~Panizzi, B.~Proiet\-ti, G.~Sacco and
K.-H.~Sulanke for important contributions in the early phases of
the project.

\end{document}